\documentclass[prb,showpaces,preprintnumbers,amsmath,amssymb,twocolumn,superscriptaddress]{revtex4-1}
\usepackage{graphicx}
\usepackage{dcolumn}
\usepackage{bm}
\usepackage{amssymb}
\usepackage{amsmath}
\usepackage{epstopdf}
\usepackage{booktabs}
\usepackage{threeparttable}
\usepackage{color}
\usepackage[pdfstartview=FitH, CJKbookmarks=true, bookmarksnumbered=true, bookmarksopen=true, colorlinks, pdfborder=001, linkcolor=blue, anchorcolor=blue, citecolor=blue]{hyperref}

\begin{document}

\title{Probing the superconducting gap structure of (Li$_{1-x}$Fe$_x$)OHFeSe}
\author{M. Smidman}
\affiliation{Center for Correlated Matter and Department of Physics, Zhejiang University, Hangzhou 310058, China}
\author{G. M. Pang}
\affiliation{Center for Correlated Matter and Department of Physics, Zhejiang University, Hangzhou 310058, China}
\author{H. X. Zhou}
\affiliation{Beijing National Laboratory for Condensed Matter Physics, Institute of Physics, Chinese Academy of Sciences, Beijing 100190, China}
\author{N. Z. Wang}
\affiliation{Hefei National Laboratory for Physical Science at Microscale and Department of Physics, University of Science and Technology of China, Hefei, Anhui 230026, China}
\affiliation{Key Laboratory of Strongly Coupled Quantum Matter Physics, University of Science and Technology of China, Hefei, Anhui 230026, China}
\author{W. Xie}
\affiliation{Center for Correlated Matter and Department of Physics, Zhejiang University, Hangzhou 310058, China}
\author{Z. F. Weng}
\affiliation{Center for Correlated Matter and Department of Physics, Zhejiang University, Hangzhou 310058, China}
\author{Y. Chen}
\affiliation{Center for Correlated Matter and Department of Physics, Zhejiang University, Hangzhou 310058, China}
\author{X. L. Dong}
\affiliation{Beijing National Laboratory for Condensed Matter Physics, Institute of Physics, Chinese Academy of Sciences, Beijing 100190, China}
\affiliation{University of Chinese Academy of Sciences, Beijing 100049, China}
\author{X. H. Chen}
\affiliation{Hefei National Laboratory for Physical Science at Microscale and Department of Physics, University of Science and Technology of China, Hefei, Anhui 230026, China}
\affiliation{Key Laboratory of Strongly Coupled Quantum Matter Physics, University of Science and Technology of China, Hefei, Anhui 230026, China}
\affiliation{Collaborative Innovation Center of Advanced Microstructures, Nanjing 210093, China}
\author{Z. X. Zhao}
\affiliation{Beijing National Laboratory for Condensed Matter Physics, Institute of Physics, Chinese Academy of Sciences, Beijing 100190, China}
\affiliation{University of Chinese Academy of Sciences, Beijing 100049, China}
\author{H. Q. Yuan}
\email{hqyuan@zju.edu.cn}
\affiliation{Center for Correlated Matter and Department of Physics, Zhejiang University, Hangzhou 310058, China}
\affiliation{Collaborative Innovation Center of Advanced Microstructures, Nanjing 210093, China}

\date{\today}

\begin{abstract}
We report measurements of the London 
penetration depth [$\Delta\lambda(T)$] of the recently discovered iron-based superconductor (Li$_{1-x}$Fe$_x$)OHFeSe, in order to characterize the nature of the superconducting gap structure. At low temperatures, $\Delta\lambda(T)$ displays nearly temperature independent behavior, indicating a fully open superconducting gap. We also analyze the superfluid density  $\rho_s(T)$ which cannot be well accounted for by a single-gap isotropic $s$-wave model but are consistent with either two-gaps, a model for the orbital selective $s\times\tau_3$ state or anisotropic $s$-wave superconductivity.

\end{abstract}

\maketitle
\section{introduction}
Iron selenide based superconductors have come to attract particular attention after the discovery that the binary material FeSe avoids the formation of the magnetic order but becomes  superconducting at $T_c$~=~8~K. \cite{hsu2008superconductivity} The value of $T_c$ can  be enhanced both by applying pressure, reaching a maximum value of about 37~K,\cite{medvedev2009electronic,margadonna2009pressure} or in  single layer films of FeSe grown on an SrTiO$_3$ substrate, \cite{qing2012interface,FeSe2013,FeSe2015} and it is suggested that large enhancements of $T_c$ arise due to an increased  electron carrier density. \cite{FeSeCarr} Another method of enhancing the superconductivity is by  intercalating alkali metals between  FeSe layers, as in the case of $A_x$Fe$_{2-y}$Se$_2$ ($A$~=~K, Rb, Cs,Tl/Rb, Tl/K) \cite{guo2010superconductivity,2011CsSCRep,wang2011superconductivity,2011TlRbSCRep,2011TlKSCRep}. A notable difference from both the iron arsenide based materials and the bulk binary compound FeSe is the  absence of the hole pocket in the Fermi surface at the Brillouin zone center.\cite{qian2011absence} This appears to contradict the  s$_{\pm}$ superconducting state often applied to iron arsenide superconductors, where there is nesting and a sign change of the order parameter between the hole and electron pockets.\cite{mazin2008unconventional,kuroki2008unconventional} While there have been a variety of alternative proposals for the pairing state, \cite{mazin2011symmetry,khodas2012interpocket,nica2015orbital} studying the intrinsic superconducting properties is greatly complicated by the clear evidence for phase separation between non-superconducting regions with an ordered arrangement of Fe vacancies and vacancy free superconducting regions. \cite{li2012phase,PhaseSep2011DL,PhaseSepNMR}.

Recently a new iron selenide based superconductor   (Li$_{1-x}$Fe$_x$)OHFeSe ($x\approx0.2$) was discovered, \cite{lu2015coexistence} with a high transition temperature of $T_c\approx$~40~K. It has a  quasi-two-dimensional crystal structure, with layers  of both (Li$_{1-x}$Fe$_x$)OH and superconducting FeSe. The   material displays   coexistence between superconductivity and  antiferromagnetism, \cite{lu2015coexistence} while in the phase diagram superconductivity occurs in close proximity to spin-density wave order. \cite{dong2014phase}  In common with $A_x$Fe$_{2-y}$Se$_2$, the hole pocket is also absent,\cite{niu2015surface,zhao2016common} but  the samples are much more homogeneous, indicating that (Li$_{1-x}$Fe$_x$)OHFeSe is a good candidate for probing iron based superconductors without a hole pocket at the zone center. Angle resolved photoemission spectroscopy (ARPES) measurements are consistent with the presence of nodeless superconductivity with a single isotropic energy gap, but  disagree over the gap magnitudes.\cite{niu2015surface,zhao2016common} However,  scanning tunneling spectroscopy (STS) studies show two distinct features in the conductance spectra, suggesting the presence of multiple gaps. \cite{du2016scrutinizing,yan2016surface} Meanwhile the in-plane superfluid density obtained from muon-spin rotation ($\mu$SR) measurements is consistent with either one or two gaps, but very different behavior is seen in the out-of-plane component, which shows a much more rapid drop of the superfluid density with temperature. \cite{khasanov2016proximity} Furthermore,  inelastic neutron scattering (INS) measurements give evidence for the presence of a spin resonance peak, \cite{Davies2016spin,pan2016structure} consistent with a sign change of the order parameter across the Fermi surface, while a lack of a sign change was suggested from an STS study, on the basis  of quasi-particle interference (QPI) results and the effect of impurities.\cite{yan2016surface}

To further characterize the superconducting gap structure,  we report London penetration depth measurements of (Li$_{1-x}$Fe$_x$)OHFeSe single crystals  using a tunnel-diode-oscillator (TDO) based technique, from which we obtain the temperature dependence of the London penetration depth shift $\Delta\lambda(T)$. The low temperature $\Delta\lambda(T)$ gives clear evidence for nodeless superconductivity, while a single-gap isotropic $s$-wave model is unable to account for the superfluid density.  The superfluid density is well fitted by a two-gap $s$-wave model, as well as models with anisotropic gaps.

\begin{figure}[t]
\begin{center}
  \includegraphics[width=\columnwidth]{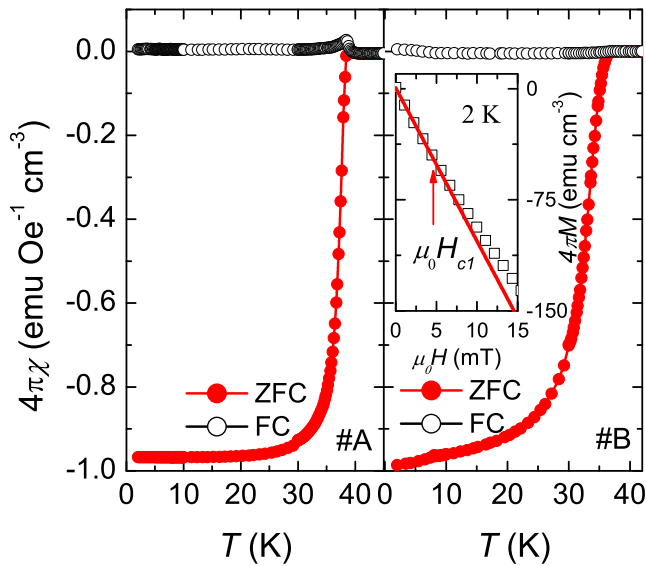}
\end{center}
	\caption{Magnetic susceptibility $\chi$ of  (Li$_{1-x}$Fe$_x$)OHFeSe samples from two batches ($\#$A and $\#$B). Both field-cooled (FC) and zero-field-cooled (ZFC) measurements were performed in an applied magnetic field of 1~mT. The inset shows the magnetization as a function of applied field at 2~K. The solid line shows the linear fit at low fields while the arrow marks the position of the lower critical field $H_{c1}$, where there is deviation from linear behavior.}
   \label{samplecharacterization}
\end{figure}
\section{Experimental details}
Single crystals of (Li$_{1-x}$Fe$_x$)OHFeSe (with $x\approx0.2$) from two batches prepared by two different groups were measured ($\#$A and $\#$B), where the crystals were synthesized following Ref.~\onlinecite{DongSynth}. Using   the parameters from   Ref.~\onlinecite{DongSynth} ($\rho_0=0.1~$m$\Omega$-cm and a carrier density of $n=1.04\times10^{21}$cm$^{-3}$), we estimate a mean free path of $l=12.4$~nm using $l=\hbar(3\pi^2)^{\frac{1}{3}}/e^2n^{\frac{2}{3}}\rho_0$. This is considerably larger than the Ginzburg-Landau coherence length $\xi_{GL}=2$~nm calculated from an upper critical field of $H_{c2}(0)=79$~T, \cite{DongSynth} and therefore the material is in the clean limit. Magnetization measurements of samples from both batches were performed utilizing a SQUID magnetometer (MPMS-5T). The temperature dependence of the London penetration depth shift $\Delta\lambda(T)~=~\lambda(T)-\lambda(0)$ was measured in a ${^3}$He cryostat from 42~K down to a base temperature of about 0.5~K using a tunnel-diode-oscillator based method, with an operating frequency of about 7~MHz. The samples were cut into a regular shape and mounted on a sapphire rod which was inserted into the coil without any contact. A very small ac field of about 2~$\mu T$ is applied to the sample along the $c$~axis, which is much smaller than the lower critical field $\mu_0H_{c1}$ and therefore the sample is always in the Meissner state. As such the   shift in the resonant frequency from zero temperature $\Delta f(T)$ is related to the penetration depth shift in the $ab$~plane $\Delta\lambda(T)$ via $\Delta\lambda(T)$~=~G$\Delta f(T)$, where the calibration constant $G$  is calculated using the geometry of the coil and sample \cite{Gfactor}. 

\begin{figure}[t]
\begin{center}
  \includegraphics[width=0.8\columnwidth]{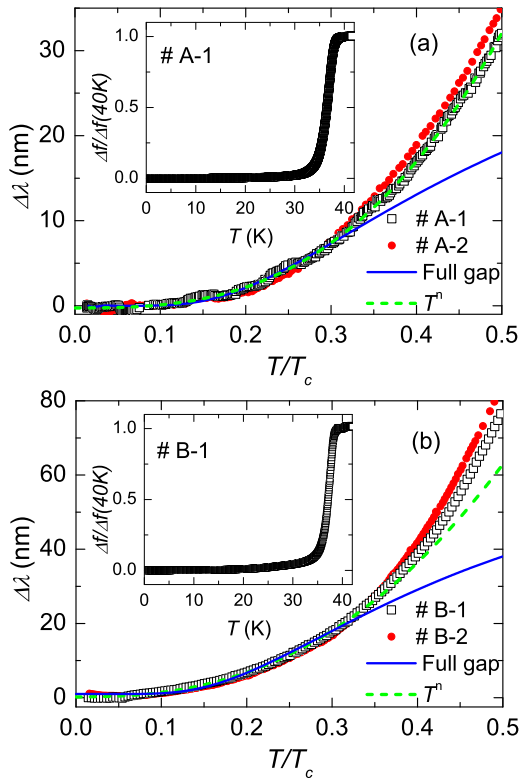}
\end{center}
	\caption{Temperature dependence of the London penetration depth $\Delta\lambda(T)$ of two samples from (a) batch $\#$A and (b) batch $\#$B. The main panels show the low temperature behavior, where the data are fitted below $T_c/3$ with  a fully gapped model and a power law dependence. The insets display the frequency shift $\Delta f(T)$ from the lowest temperature up to above the superconducting transition temperature, normalized by the value at 40~K.}
   \label{penetrationdepth}
\end{figure}

\section{Results and discussion}

 Both the field-cooled (FC) and zero-field-cooled (ZFC) magnetic susceptibility [$\chi(T)$] measurements are shown in Fig.~\ref{samplecharacterization}, from above the superconducting transition temperature down to 2~K under a small applied field of 10~Oe, where corrections for the demagnetization effect were applied. The ZFC $\chi(T)$ of both samples show sharp superconducting transitions onsetting at  around 40~K
and 37~K for samples $\#$A and $\#$B, respectively. At low temperatures the data for $\#$A  flattens, while
for  $\#$B there continues to be a slight decrease with decreasing temperature. This indicates the high quality of the single crystals from batch $\#$A, whereas those from batch $\#$B show evidence for some inhomogeneity. Meanwhile the superconducting shielding fraction is around 100$\%$ in both samples. The inset displays the field dependence of the magnetization at 2~K, for a field applied in-plane so that the demagnetization effect is very small. There is a deviation of the low field linear behavior at $\mu_0H_{c1}\approx 4.5$~mT, confirming that $H_{c1}$ is significantly larger than the ac field applied in the  penetration depth measurements.

Figure~\ref{penetrationdepth} displays $\Delta\lambda(T)$ for single crystal samples from two batches, with samples from $\#$A and $\#$B displayed in (a) and (b) respectively. The insets of both panels display the temperature dependence of the frequency shift $\Delta f(T)$ from above the superconducting transition at 42~K  down to 0.5~K. The superconducting transition onsets at respective temperatures of 40~K and 39~K in samples $\#$A and $\#$B, while the corresponding endpoints of the transition are around 34~K and 35~K, and the latter values of $T_c$ are used in the subsequent analysis of the superfluid density. The main panels of Fig.~\ref{penetrationdepth} display the low temperature behavior of $\Delta\lambda(T)$. It can be seen in Fig.~\ref{penetrationdepth}(a) that $\Delta\lambda(T)$ for the  $\#$A samples decreases with decreasing temperature before flattening below about 0.1T$_c$, indicating a nodeless gap structure in (Li$_{1-x}$Fe$_x$)OHFeSe with a lack of low energy excitations, which is consistent with previous results. \cite{niu2015surface,zhao2016common,du2016scrutinizing,yan2016surface,khasanov2016proximity} Furthermore when fitted with a power law dependence  $\Delta\lambda(T)\sim T^n$, exponents of $n=2.83$ and $n=2.46$ are obtained for samples  $\#$A-1 and  $\#$B-1 respectively. In the case of nodal superconductors, $n=1$ for line nodes and $n=2$ for point nodes is generally anticipated. While impurity scattering, non-local effects and quantum fluctuations can all lead to  $n\approx2$ at low temperatures for $d$-wave superconductors with line nodes,\cite{Hirschfeld1993,Kosztin1997,Benfatto2001} our observation that $n$ is significantly larger than two gives further evidence for fully gapped behavior. For a fully-gapped superconductor at $T\ll T_c$, the penetration depth can be described by $\Delta\lambda(T)=\lambda_{eff}(0)\sqrt{\pi\Delta(0)/2k_BT}\textrm{exp}[-\frac{\Delta(0)}{k_BT}]$, where $\Delta(0)$ is the superconducting gap magnitude at zero temperature and $\lambda_{eff}(0)$ is an effective zero temperature penetration depth. The low temperature data  for sample $\#$A-1 is well fitted with a gap magnitude of  $\Delta(0)=~0.87k_BT_c$. A similar  value of $\Delta(0)=~0.78k_BT_c$ is obtained from the fitting for  sample $\#$B-1, although there is a small deviation of the  fitted curve for this sample at the lowest temperatures, where a weak anomaly is observed in the data, the origin of which  is not clear. The values of the fitted gaps are much smaller than the value from BCS theory of $1.76k_BT_c$ for weakly coupled isotropic $s$-wave superconductors, suggesting  multi-gap superconductivity or gap anisotropy in (Li$_{1-x}$Fe$_x$)OHFeSe. The fitted values of  $\lambda_{eff}(0)$ of 636~\AA\ and 1228~\AA\ for samples  $\#$A-1 and  $\#$B-1 respectively, are significantly different from the value of $\lambda(0)\approx2800$~\AA\ from  $\mu$SR measurements. \cite{khasanov2016proximity}. Such a difference is also expected for multi-gap or anisotropic superconductors. \cite{Malone2009}

\begin{figure}[t]
\begin{center}
  \includegraphics[width=\columnwidth]{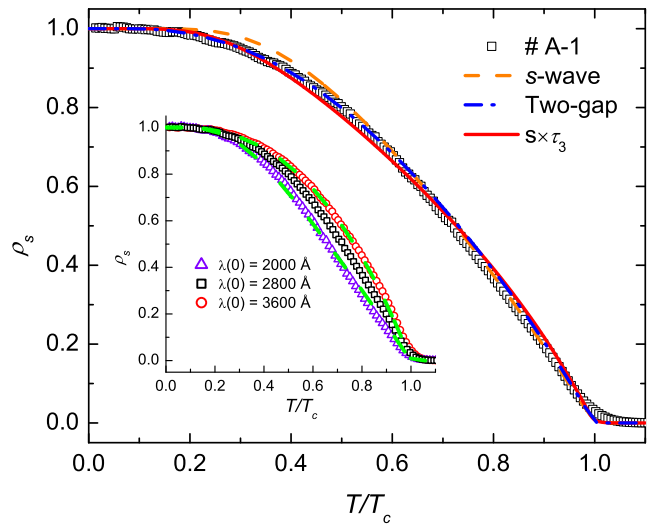}
\end{center}
	\caption{Normalized superfluid density $\rho_s$ as a function of the reduced temperature $T/T_c$ for sample $\#$A-1 of (Li$_{1-x}$Fe$_x$)OHFeSe. The dashed, dashed-dotted and solid lines show fits to models with a single isotropic $s$-wave gap, two isotropic gaps and the $s\times\tau_3$ state respectively.   The inset shows  $\rho_s$ upon varying $\lambda(0)$ by $\pm30\%$, along with fits to a two-gap model.}
   \label{superfluiddensity}
\end{figure}

In order to obtain further information about the superconducting pairing state, the normalized superfluid density [$\rho_s(T)$] was calculated  from the penetration depth using  $\rho_s(T)$~=~$[\lambda(0)$/$\lambda(T)]^2$, where $\lambda(0)\approx$~2800~\AA\ was estimated from  $\mu$SR measurements.\cite{khasanov2016proximity} Since the samples from batch  $\#$A are  higher quality and show a sharper superconducting transition, measurements from this batch were used in the subsequent analysis. The $\rho_s(T)$ of sample $\#$A-1 is displayed in Fig.~\ref{superfluiddensity}.  The superfluid density was modelled using 
\begin{equation}
\rho_{\rm s}(T) = 1 + 2 \left\langle\int_{\Delta_k}^{\infty}\frac{E{\rm d}E}{\sqrt{E^2-\Delta_k^2}}\frac{\partial f}{\partial E}\right\rangle_{\rm FS},
\label{equation2}
\end{equation}
\noindent where $f(E, T)=[1+{\rm exp}(E/k_BT)]^{-1}$ is the Fermi function and $\left\langle\ldots\right\rangle_{\rm FS}$ denotes an average over the Fermi surface. The temperature dependence of the gap function $\Delta_k$ is approximated by \cite{carrington2003magnetic}
\begin{equation}
\delta(T)={\rm tanh}\left\{1.82\left[1.018\left(T_c/T-1\right)\right]^{0.51}\right\},
\label{equation3}
\end{equation}

\noindent As discussed previously, the behavior of $\Delta\lambda(T)$ at low temperatures and previous experimental results indicate nodeless superconductivity in (Li$_{1-x}$Fe$_x$)OHFeSe and therefore we fitted $\rho_s(T)$ with various fully-gapped models. The simplest model is to assume an isotropic superconducting gap, with $\Delta_k(T)~=~\Delta(0)\delta(T)$. The fitted curve for such an isotropic single band $s$-wave model with $\Delta(0)~=~1.72k_BT_c$ is  shown by the dashed line in Fig.~\ref{superfluiddensity}, where although there is reasonable agreement above 0.5$T_c$, there is a significant discrepancy in the intermediate temperature range between 0.2$T_c$ and 0.5$T_c$. This difference arises due to the data dropping more quickly than the calculated $\rho_s(T)$, indicating the presence of an additional lower energy scale in the gap structure, which is consistent with the smaller gap value obtained from fitting  $\Delta\lambda(T)$ at low temperatures. Therefore the data are fitted with a two-gap model, as has been applied to many iron based superconductors. For this model the total superfluid density is given by the weighted sum of two components with different gaps,
\begin{equation}
\rho_{\rm s}(T) = \alpha\rho{_{\rm s}^1}(\Delta_k^1, T) + (1-\alpha)\rho{_{\rm s}^2}(\Delta_k^2, T),
\label{equation4}
\end{equation}

\noindent where $\rho_s^i (i=1, 2)$ is the normalized superfluid density with a gap function $\Delta_k^i (i=1, 2)$  and $\alpha$ is the relative weight for the component $\rho_s^1$ ($0\leq\alpha\leq1$).  The fitting result is shown in Fig.~\ref{superfluiddensity}  by the dashed-dotted  line, which agrees well with the data across the whole temperature range.  The fitted parameters are $\Delta_1(0)=0.8k_BT_c$, $\Delta_2(0)=1.9k_BT_c$ and $\alpha=0.13$, where the value of the small gap is close to that found from fitting $\Delta\lambda(T)$, which is further consistent with two-gap superconductivity. In order to consider possible uncertainties in the calibration constant $G$ or $\lambda(0)$, in the inset we show $\rho_s(T)$  upon varying $\lambda(0)$ by $\pm30\%$. The data can still be fitted by a two gap model with slightly different parameters of $\Delta_1(0)=0.8k_BT_c$, $\Delta_2(0)=1.6k_BT_c$ and $\alpha=0.15$ for  $\lambda(0)$~=~2000~\AA ~and $\Delta_1(0)=0.8k_BT_c$, $\Delta_2(0)=2.1k_BT_c$ and $\alpha=0.12$ for  $\lambda(0)$~=~3600~\AA. The data were also well fitted with an anisotropic single band model with $\Delta_k(T,\phi)=\Delta(0)(1+r{\rm cos}2\phi)\delta(T)$, using $\Delta(0)=1.32k_BT_c$ and $r=0.65$, which is not shown for the sake of clarity.

ARPES measurements indicate that the Fermi surface consists of electron pockets at the Brillouin zone corners, without the presence of hole pockets at the zone center \cite{niu2015surface,zhao2016common}. From a recent STM study it was proposed on the basis of QPI measurements, as well as examining the effects of impurities, that there is no sign change of the superconducting gap across the Fermi surface \cite{yan2016surface}, in which case such a two gap $s$-wave model readily explains the data. However INS measurements show evidence for a spin resonance peak,\cite{Davies2016spin,pan2016structure} which indicates that there is a sign change of the order parameter between regions of the Fermi surface connected by the resonance wave vector. This would be incompatible with two-gap $s$-wave superconductivity with no sign change and are also difficult to account for with the $s_{\pm}$ state proposed for  many iron arsenide superconductors, due to both the lack of a hole pocket at the zone center and a different nesting wave vector for the spin resonance. \cite{mazin2008unconventional,kuroki2008unconventional} On the other hand a different sign changing $s_{\pm}$ state has been suggested for $A_x$Fe$_{2-y}$Se$_2$  (A~=~K, Rb, Cs), \cite{khodas2012interpocket,mazin2011symmetry} with a very similar Fermi surface to  (Li$_{1-x}$Fe$_x$)OHFeSe.

Another proposed pairing state for nodeless sign changing superconductivity in iron based superconductors with only electron pockets is an orbital selective $s\times\tau_3$ state. \cite{nica2015orbital} In this scenario, intraband pairing has $d_{x^2-y^2}$ symmetry,  while the  interband pairing has $d_{xy}$ symmetry. As a result, the zeroes of the intraband and interband pairing  are offset by an angle of $\pi/4$ and therefore the gap remains nodeless. A simple model for the gap function of this state taking into account the Fermi surface is  $\Delta_k(T,\phi)$=[($\Delta_1(0)$)$^2$+($\Delta_2(0)$sin($\phi$))$^2$]$^{1/2}$$\delta(T)$ \cite{nica2015orbital,SiPrivate}. As shown in Fig.~\ref{superfluiddensity}, this model can also well fit the experimental data, with fitted parameters of $\Delta_1(0)=1.05k_BT_c$ and $\Delta_2(0)=3.2k_BT_c$. In this case the gap minimum $\Delta_1(0)$ is slightly larger than the smaller gap from the two-gap $s$-wave fit. Therefore the data can be well accounted for by fitting with models with either two-gaps or a strong gap anisotropy and our results are compatible with two-gap behavior, anisotropic $s$-wave superconductivity or an orbital selective $s\times\tau_3$ state. It is often very difficult to distinguish between scenarios with multiple gaps and those where there is one anisotropic gap. While the isotropic nature of the gap inferred from ARPES would favor the two-gap scenario over an anisotropic gap, this still requires further study. \cite{niu2015surface,zhao2016common} Furthermore, the superfluid density is only sensitive to the gap magnitude rather than the phase and  different measurements are required to clarify the presence of a sign change  and to determine the nature of the pairing state.

We note that there have been conflicting reports about the nature of the gap structure of (Li$_{1-x}$Fe$_x$)OHFeSe, with only a single gap being resolved from ARPES measurements \cite{niu2015surface,zhao2016common}, while two gaps are found from STS \cite{du2016scrutinizing,yan2016surface} and the in-plane superfluid density from $\mu$SR  measurements is compatible with both single-gap and two-gap models. \cite{khasanov2016proximity} The gap values we obtain from fitting the in-plane superfluid density are smaller than those reported from previous measurements, particularly in the case of the smaller gap. Evidence for this smaller gap is clearly observed from our measurements of $\Delta\lambda(T)$ and $\rho_s$ at low temperatures, which may be a result of the high sensitivity of the TDO-based technique. A further reason for the discrepancies could be due to the non-stoichiometric nature of (Li$_{1-x}$Fe$_x$)OHFeSe, where the exact composition and homogeneity may influence the doping level, $T_c$ and the gap magnitudes. In addition, the doping level can be different between the surface and in the bulk, \cite{niu2015surface} and therefore probes which primarily measure the surface properties may give different results. Indeed different results have been found from different measurements of the gap structure of other iron-based superconductors, \cite{evtushinsky2009momentum} and significant sample dependence has been suggested for Ba$_{1-x}$K$_x$Fe$_2$As$_2$.\cite{hashimoto2009microwave}

\section{summary}

We have measured the temperature dependence of the London penetration depth $\Delta\lambda(T)$ and the derived superfluid density $\rho_s(T)$ of the recently discovered high-$T_c$ iron-based superconductor (Li$_{1-x}$Fe$_x$)OHFeSe. The behavior of $\Delta\lambda(T)$ at low temperatures gives clear evidence for nodeless superconductivity, with a relatively small energy gap, while the analysis of $\rho_s(T)$ is consistent with both two-gap superconductivity and models with significant gap anisotropy such as an orbital selective $s\times\tau_3$ or anisotropic $s$-wave  state.

We thank Q. Si, E. M. Nica, R. Yu and X. Lu for helpful discussions and valuable suggestions. This work was supported by the National Natural Science Foundation of China (No.11574370), National Key Research and Development Program of China (No. 2016YFA0300202), and the Science Challenge Project of China.

\end{document}